\documentclass{aa}
\usepackage{graphicx}
\usepackage{amssymb}
\def\K{K$\rm _s$~}

\begin{document}
   \title{NOVA Sco 2001 (V1178 SCO)
\thanks{Based on observations made with the
Danish 1.5-m \protect{\newline}telescope and the 1.0-m telescope at ESO, La Silla Chile}}

\author{
M. Andersen\inst{1} \and
S. Kimeswenger\inst{2}
}

\offprints{S. Kimeswenger,\\~~\email{stefan.kimeswenger@uibk.ac.at}}

\institute{
Astrophysikalisches Institut Potsdam, An der Sternwarte 16, D-14482 Potsdam, Germany \and
Institut f{\"u}r Astrophysik der Universit{\"a}t Innsbruck, Technikerstr. 25,
A-6020 Innsbruck, Austria
}

   \date{Received 29. June2001 / accepted 13. August 2001}

\abstract{
We present intermediate resolution spectroscopy and near infrared photometry of NOVA Sco 2001 
 (V1178 Sco), which was first detected May 13$^{\rm th}$ 2001 and reported June 21$^{\rm th}$ 2001,
and obtained by us the same day.  We also retrieved very accurate astrometry of the target in 
this very crowded field.
This is needed to
be able to do follow up observations of the postnova during the next years.
The spectrum shows an overall expansion of 2100~km~s$^{-1}$ and
has clearly complex, and most likely nonsymmetric, outflow
substructures.
We clearly identify this object as classical nova, "Fe II" subclass. 
}
\titlerunning{The NOVA Sco 2001 (V1178 SCO)}
\maketitle


\section{Introduction}
NOVA Sco 2001 was discovered May 13$^{\rm th}$ 2001 as a 10$\,.\!\!^{\rm \tt m}$5 object
and reported first June 21$^{\rm th}$ 2001, 18:03 UT (Hasada et al. \cite{IAUC_A}). 
Liller (\cite{IAUC_B})
reports there also first spectroscopic results at low dispersion, confirming a strong
H$_\alpha$ line. The variable star name given to the object is V1178 Sco (Samus \& Kazarovets 
\cite{IAUC_D}).
We obtained near infrared images in the Gunn-i, J and K$_{\rm s}$ band
using the DENIS survey instrument (Epchtein et al. \cite{Epchtein})
attached to the ESO/LaSilla 1m telescope using
5$\,.\!\!^{\rm \tt m}$2 attenuating filters (June 21$^{\rm th}$ 2001, 23:12 UT).
The spectrum was taken using the
Danish 1.5m telescope at LaSilla (June 22$^{\rm th}$ 2001, 02:46 UT)
with the DFOSC mounted.
Also a red continuum filter image was obtained there.
This filter was selected to have a largely line emission free narrow band
image, to get best possible astrometry without effects of color shifts and differential refraction
of the surrounding astrometric calibration sources. 

\section{Astrometry and Cross--identification}

To obtain a very accurate astrometry of the target a DFOSC narrow band image was taken.
To avoid displacements of sources due to different colors as seen 
normally in wide band filters, the red continuum filter ESO\#840 centered
at 683.82~nm and having a FWHM of 8.09~nm was used. The DFOSC is currently
equipped with a MAT/EEV 44-82 2k$\times$4k CCD giving, according to the
manual, a resolution of 0\farcs39 per pixel. In fact we measure 0\farcs39556$\pm$0\farcs00018.
The FWHM on the images was 1\farcs66. As only half the chip is used the field of view 
is 13$\times$13\arcmin.
We used only the central part of the image with a radius of 1\arcmin~ around
the target. This provides us with a highly distortion free image. 
Astrometric calibrators were taken from {\it USNO CCD Astrometric Catalogue} (UCAC) (Zacharias et al.
\cite{UCAC}). This catalogue contains southern sources with an accuracy of about 20~mas
in the red magnitude range $10^{\rm \tt m} < {\rm m} < 14^{\rm \tt m}$ and still has  
an accuracy of about 70~mas at the limit of 16$^{\rm \tt m}$.
We used 6 stars surrounding the target to obtain the astrometry (Tab. \ref{astro}). 
One source corresponds to a TYCHO-2 source. The difference of the position in the
2 catalogues is, after correction for proper motion, approximately 25~mas in both coordinates. 
\begin{table}
[h]
\caption{Astrometric calibration stars}
\label{astro}
\begin{tabular}{l l l l}
\hline
UCAC Id. &  TYCHO-2 Id. & $\alpha$ (J2000.0) &  $\delta$ (J2000.0) \\
\hline\hline
19383082 & 1837251 & 269.3048774 & -32.4037892 \\
19383048 &         & 269.2938362 & -32.3957092 \\
19383013 &         & 269.2828100 & -32.3929684 \\
19383009 &         & 269.2820474 & -32.3899953 \\
19382967 &         & 269.2719486 & -32.3779859 \\
19382986 &         & 269.2773877 & -32.3915264 \\
\hline
\end{tabular}
\vspace{-5mm}
\end{table}
\begin{figure}[ht]
\resizebox{\hsize}{!}{\phantom{XXXXXXXXXXXXXXXXXXXX}\includegraphics{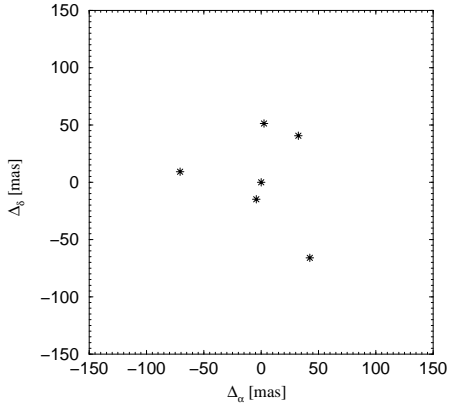}\phantom{XXXXXXXXXXXXXXXXXXXX}}
\caption{The scatter diagram of the astrometric calibration sources around the target.}
\label{astro_fig}
\vspace{-3mm}
\end{figure} 

The source extraction on the image was obtained by using SExtractor v2.1.6 
(Bertin \& Arnouts \cite{Bertin}). The rms of the positions, using a plane
astrometry without distortion coefficients, was 41~mas.
The 2 largest residuals (up to 71
mas) are found for the two faintest sources. As our S/N was very high even for
those sources, we assume that part of this error originates from the UCAC.
These residuals are in the order of magnitude for the input catalogue itself. 
The 1$\times$1\arcmin\ region
was thus sufficent.
 It would give us just uncertainties due to distortion increasing
the field.
As the target is very bright, we assume the accuracy of our coordinates 
to be 
40~mas.:

\smallskip
\centerline{
$\alpha_{\rm \tt J2000.0} = 17^{\rm \tt h}57^{\rm \tt m}06\,.\!\!^{\rm \tt s}922\, \pm 0\,.\!\!^{\rm s}003$}
\centerline{$\delta_{\rm \tt J2000.0} = -32\degr25'05\farcs03\, \pm 0\farcs04$~.}

An inspection of the sky survey plates SERC V (1987.590 and 1987.708) and
2$^{\rm nd}$~ed.~Equatorial~red (1989.674 and 1992.421) shows a source, which is 
significantly fainter than the red 17\fm5 source stated in Hasada et al. (\cite{IAUC_A}).
Although this source is faint, we are able to state that there is no variability at a level of 0\fm25 at those
epochs. The CAI MAMA scans provided at CDS give an estimate of B$\leq$20\fm5 and R$\leq$18\fm2 when
using the PSF fitting for decrowding and photometric calibration as described in Kimeswenger \& Weinberger (\cite{KW01}).
The DENIS \K ~images were used together with the accurate astrometry above, to
carefully cross--identify the 2MASS sources of the region. We thus are able to definitely exclude
the next nearby source 1757066-322305 as possible progenitor.
This source is  westward to the target still visible on the \K~images.
\begin{figure}[ht]
\resizebox{\hsize}{!}{
\phantom{XXXXXXXXXXXX}
\includegraphics{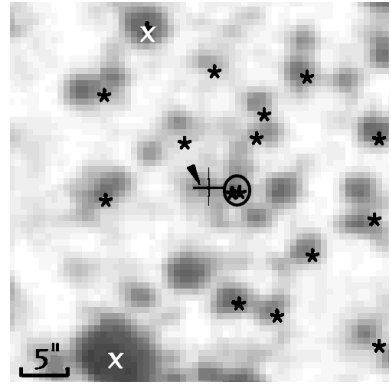}
\phantom{XXXXXXXXXXXX}
}
\caption{The finding chart for the target from the CDS/Aladin MAMA CAI red scan. 
The coordinates from the astrometry are centered (fineline cross). We assume that the faint star marked
with the arrow is the real progenitor. The UNSO A2 sources (asterix) and the UCAC sources ($\times$)
are marked. The USNO sources 0523-30816527 and 0525-30186320, being identical with the 2MASS source 
1757066-322305 (circle), are definitely not the progenitor(s) (see text).}
\label{image}
\vspace{-3mm}
\end{figure} 
Fig.~\ref{image} shows, that the two USNO A2 sources mentioned by Masi (\cite{IAUC_C}) 
 (0523-30816527 and 0525-30186320, one star in fact only)
are
the same source as the 2MASS source mentioned above. It has a significant offset to the
target. The coordinate offset of Masi may originate from the fact that
this target, observed from Italy, had a high airmass. He underestimated the effects of
differential refraction when comparing his unfiltered CCD image with the USNO red.
Queries to different public archives at e.g. ESO, ING, etc. gave no results concerning an
earlier CCD image.
\section{NIR Photometry}

The near infrared photometry was obtained at the DENIS survey instrument
(Epchtein et al. \cite{Epchtein})
attached to the ESO/LaSilla 1m telescope. We used a Gunn-i ($\lambda_{\rm eff}$~=~0.81~$\mu$m)
a standard J ($\lambda_{\rm eff}$~=~1.25~$\mu$m) and a \K ($\lambda_{\rm eff}$~=~2.15~$\mu$m)
filter. The night was photometric. The standard observations were taken from the
survey run. The calibration was not done via the survey pipeline but manually.
The methods are similar to those described in Cioni et al. (\cite{Cioni}) and
Fouqu{\'e} et al. (\cite{Fouque}).
 As the source was just below overexposure in normal survey mode, also images
using a 
5$\,.\!\!^{\rm \tt m}$2 attenuating filter set were obtained. The calibration 
of those sets is described in Kimeswenger et al. (\cite{Kim01}).
Thus in total 10 frames, taking the source at different positions of the 
detector, were
obtained. This was done to minimize effects due to flatfield features and detector flaws.

\begin{table}[h]
\caption{The results of the NIR photometry}
\label{NIR}
\begin{tabular}{lcll}
Date & Band & [mag] & error \\
\hline\hline
June, 21$^{\rm th}$ 2001, 23:12 UT & Gunn-i & 9\fm75 & 0\fm03 \\
& J & 8\fm67 & 0\fm05 \\
& \K & 8\fm01 & 0\fm04 \\
\hline
\end{tabular}
\end{table}

The colours (i-J = 1\fm08; J-\K = 0\fm66) are completely different from those
derived for e.g. CI Aql during the 2000 outburst at a 
date of about 30 to 40 days after discovery using the same instrument.  
While the i-J is significantly redder than the value obtained there,
the J-\K is significantly bluer. This may be an indicator of a less dominant Br$_\gamma$
emission.

\section{Spectroscopy}

The spectra were obtained at the Danish 1.5m telescope at ESO LaSilla/Chile with the
DFOSC spectrograph.
\begin{figure*}[ht]
\resizebox{\hsize}{!}{\includegraphics{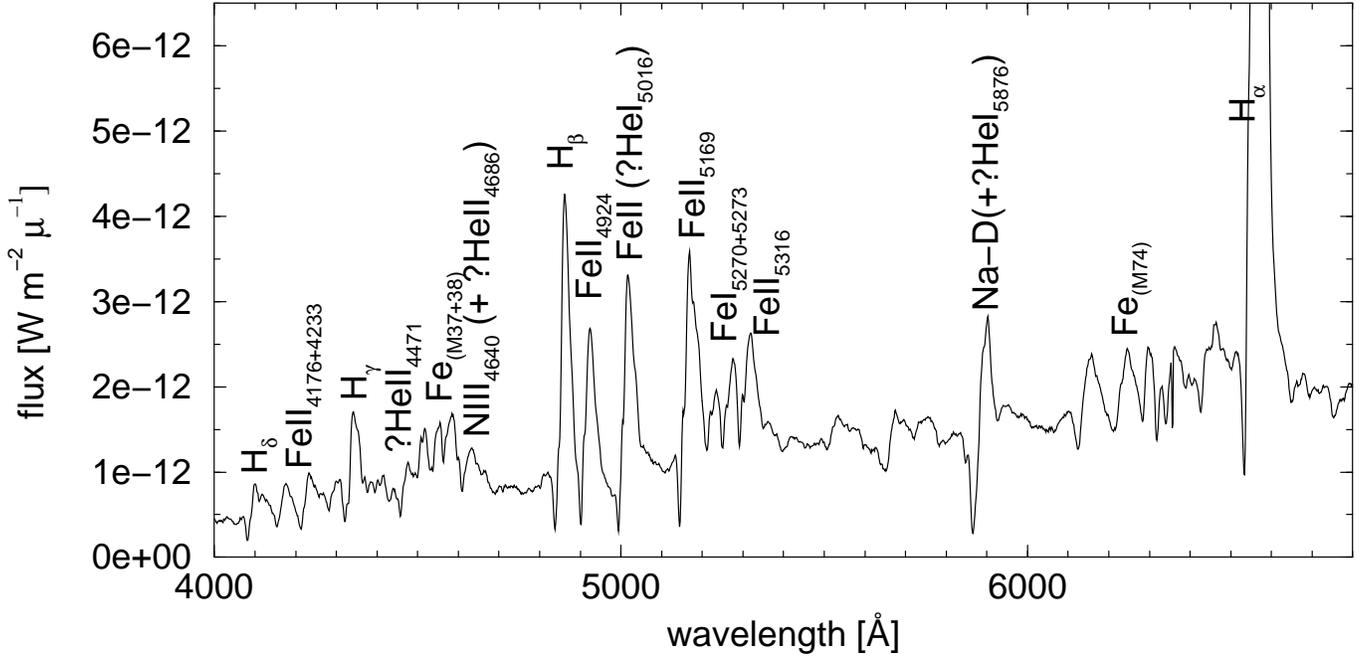}
}
\caption{The spectrum of V1178 SCO. H$_{\alpha}$ is displaced exceeding the box in order to see all
 the weaker features.
The peak is 2.8\,\,10$^{-11}$\,W\,m$^{-2}$\,$\mu$m$^{-1}$ (12.7 $\times$ the red continuum).
All P-Cygni profiles show more or less the same substructures. The maximum expansion velocity is about 2100 km\,s$^{-1}$. The slope of the continuum is consistent with the outburst having an age of 
more than one month. 
Series of Fe II lines are identified by their multiplet numbers only.}
\label{full_spectrum}
\end{figure*} 
We were using Grism \#7, giving a resolution of 0.145~nm/pixel and a usable range from
450 to 680 nm. The calibration (bias, flatfield, wavelength calibration and response curve)
 was done using usual procedures in MIDAS and the
calibration data given in the DFOSC manual. The absolute scale was calibrated by using
the photometry of Liller (\cite{IAUC_B}) and the standard V filter curve.
The spectrum has a S/N of $>$200 in the continuum over the whole region.\\
As P-Cygni profiles we are able to identify the Balmer series from H$_\alpha$ to H$_\epsilon$,
 FeII $\lambda\lambda$4176, 4233 and 5169. The high excitation lines 
NIII $\lambda\lambda$4640/HeII $\lambda\lambda$4686 do not appear 
in this nova.
Remarkably, the iron lines are stronger than e.g reported in CI Aql (Kiss et al. \cite{Kiss}), 
while  [NII] $\lambda\lambda$5755 does not appear at all. The Na-D line is very prominent too. 
Normally blended by HeI $\lambda\lambda$5876, this line appears here only as a weak
absorption feature at the blue end. 

\begin{figure}[h]
\resizebox{\hsize}{!}{
\phantom{XXXXXXX}
\includegraphics{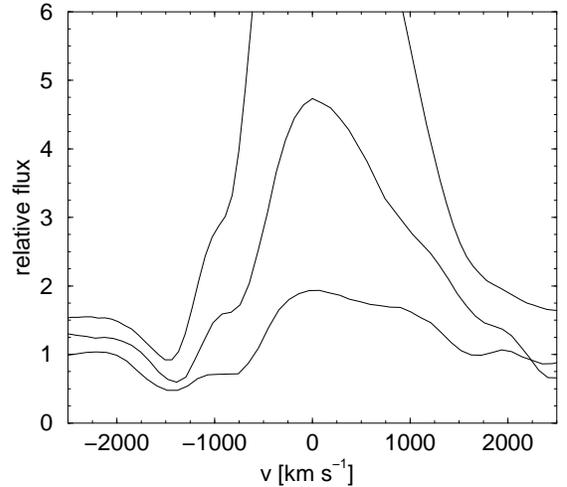}
\phantom{XXXXXXX}
}
\caption{The P-Cygni profile of the Hydrogen Balmer lines. From top to bottom: H$_\alpha$, 
H$_\beta$ and H$_\gamma$; the lines are normalized to local continuum and shifted by 0.25 each 
(see text).}
\label{balmer}
\vspace{-3mm}
\end{figure} 

The profiles of the Balmer serie lines are widely identical, providing a V$_{max}$ of 2100 km s$^{-1}$ and
two nearly symmetrical features at about 900 km s$^{-1}$. The relative strength of this feature, both
in emission as in absorption increases to higher levels. This is an indication for an inner shell not being
in thermal equilibrium.

\begin{figure}[h]
\resizebox{\hsize}{!}{
\phantom{XXXXXXX}
\includegraphics{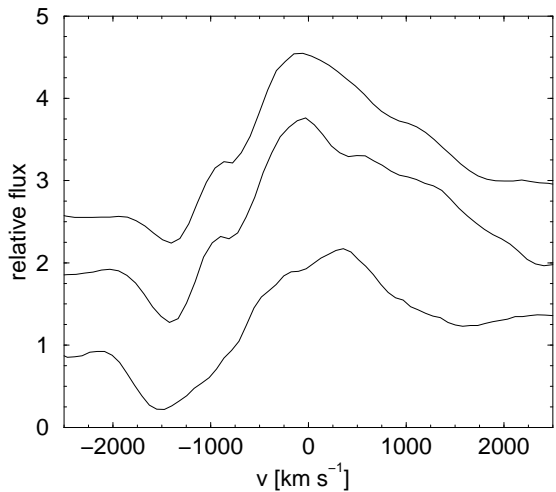}
\phantom{XXXXXXX}
}
\caption{The profiles of (from top to bottom) 
line at $\lambda\lambda$5014--5017 (Fe II$_{\rm (M42)}$, HeI ?),
FeII $\lambda\lambda$5169 and
Na-D; the lines are normalized to local continuum and shifted by 1.0 each (see text).}
\label{metal}
\vspace{-3mm}
\end{figure} 

While the structure of the line centered at 5014--5017\,\AA\ (FeII$_{\rm (M42)}$, HeI ?) follows mostly 
that of the 
hydrogen 
lines (only V$_{max}$ is lower),
the other iron and the sodium lines show much more complex features. The emission maximum is even 
suppressed by some
absorption in case of the Na-D line. 

\section{Conclusion}

The spectrum allows us to classify this object, first called as "novalike variable" in some
IAUCs, to 
be a classical nova of "Fe II" subtype 
(after Williams \cite{will92}).
The astrometry presented here clearly indicates, that the progenitors discussed
in the first IAU circulars are not the correct identifications. This will allow 
detailed followup after the decline. The outburst here, reaching $\ge$ 9$^{\rm m}$ is 
clearly stronger than that of the recent events of CI Aql or V445 Pup.
As the discovery was reported with a delay of about one month, we are unable to give
colors and accurate photometry of the early phases of the event. Thus the $t_1$ date will be poorly defined.
The blue  NIR photometry suggests $t_2$ is not reached already. This
makes it to be a moderately slow nova. 
Following the discussion in Kiss et al. (\cite{Kiss}), one may obtain 
an absolute magnitude of $-7\fm0 < M_V < -8\fm0$. 
As we have no information about the interstellar reddening
in this direction, we used the mean extinction method given for
 Miras and AGB stars in Whitelock et al. 
(\cite{WMF}). This leads to a crude distance estimate of 3 kpc and A$_V \approx 6^{\rm m}$.
 This is in agreement with the
 possible progenitor on the plates to be a late K or early M giant.\\
The spectroscopy clearly shows a complex structure of the outflow. The velocity field obtained by
the hydrogen and helium lines suggest a two shell structure similar to the models
of Hanuschik et al. (\cite{shell}). The metal lines even indicate a more complex, most likely 
non--symmetric outflow with respect to the line of sight.

\begin{acknowledgements}
We thank the referee W. Liller for his helpful suggestions.
MA is gratefull to IJAF (DK) and
KS for the BMfBWK (A) for travel support. 
\end{acknowledgements}


\begin{thebibliography}{}
\bibitem[1996]{Bertin}
Bertin, E., \& Arnouts, S., 1996, A\&AS, 117, 393

\bibitem[2000]{Cioni}
Cioni, M.-R., Loup, C., Habing, H. J., et al., 2000, A\&AS, 144, 235

\bibitem[1997]{Epchtein}
Epchtein, N., de Batz, B., Capoani, L., et al., 1997, The Messenger, 87, 27

\bibitem[2000]{Fouque}
Fouqu{\'e}, P., Chevallier, L., Cohen, M., et al. 2000, A\&AS, 141, 313

\bibitem[1993]{shell}
Hanuschik, R.W:, Spyromilio, S., Stathakis, R., Kimeswenger, S., Gochermann, J., 
Seidensticker, K.J., Meurer, G., 1993, MNRAS, 261, 909

\bibitem[2001]{IAUC_A}
Haseda, K., Kadota, K., Yamaoka, H., Takamizawa, K., Kato, T., 2001, IAUC, 7647, 1

\bibitem[2001]{Kim01}
Kimeswenger, S., Lederle, C., Armsdorfer, B., et al., 2001, A\&A, in preparation

\bibitem[2001]{KW01}
Kimeswenger, S., Weinberger, R., 2001, A\&A, 370, 991


\bibitem[2001]{Kiss}
Kiss, L.L., Thomson, J.R., Ogloza, W., F{\"u}r{\'e}sz, G., Szil{\'a}di, 2001, A\&A, 366, 858

\bibitem[2001]{IAUC_B}
Liller, W., 2001, IAUC, 7647, 2

\bibitem[2001]{IAUC_C}
Masi, G., 2001, IAUC, 7650, 2

\bibitem[2001]{IAUC_D}
Samus, N.N., Kazarovets, E., 2001, IAUC, 7650, 1

\bibitem[2000]{Schmeja}
Schmeja, S., Armsdorfer, B., Kimeswenger, S., 2000, IBVS, 4957

\bibitem[2000]{WMF} Whitelock, P.A., Marang, F.
 Feast, M., 2000, MNRAS, 319, 728

\bibitem[1992]{will92}
Williams, R.E., 1992, AJ, 104, 725

\bibitem[2000]{UCAC}
Zacharias, N., Urban, S.E., Zacharias, M.I., et al., 2000, AJ, 120, 2131
\end{thebibliography}
\end{document}